\documentclass[aap,MSNbibl,nameyear,dvips]{arximspdf}
\usepackage{dcolumn}
\usepackage{graphicx}
\doi{10.1214/11-AAP824} 
\volume{23}
\issue{1}
\pubyear{2013}
\firstpage{230}
\lastpage{250}

\makeatletter
\newcolumntype{d}[1]{D{.}{.}{#1}}
\newtheorem{theorem}{Theorem}
\newtheorem{lemma}{Lemma}[section]
\newproclaim{nm}{Numerical example}

\def\sqz{}
\makeatother

\begin{document}
\begin{frontmatter}

\title{Population genetics of neutral mutations in exponentially
growing cancer cell populations}
\runtitle{Genetics of exponentially growing populations}

\begin{aug}
\author[A]{\fnms{Rick} \snm{Durrett}\corref{}\thanksref{t1}\ead[label=e1]{rtd@math.duke.edu}}
\thankstext{t1}{Supported by NSF Grant DMS-10-05470 from the
probability program
and NIH Grant R01-GM096190-02.}
\runauthor{R. Durrett}
\affiliation{Duke University}
\address[A]{Department of Mathematics\\
Duke University\\
Box 90320\\
Durham, North Carolina 27708-0320\\
USA\\
\printead{e1}} 
\end{aug}

\received{\smonth{8} \syear{2011}}
\revised{\smonth{10} \syear{2011}}

%
\begin{abstract}
In order to analyze data from cancer genome sequencing projects, we
need to be able to distinguish causative, or ``driver,'' mutations from
``passenger'' mutations that have no selective effect. Toward this end,
we prove results concerning the frequency of neutural mutations in
exponentially growing multitype branching processes that have been
widely used in cancer modeling. Our results yield a simple new
population genetics result for the site frequency spectrum of a sample
from an exponentially growing population.
\end{abstract}

%

\end{frontmatter}
%

\section{Introduction}\label{sec1}

It is widely accepted that cancers result from an accumulation of
mutations that increase the fitness of tumor cells compared to the
cells that surround them. A number of studies
[Sj\"oblom et al. (\citeyear{Sjoetal06}), Wood et al. (\citeyear{Woo07}), Parsons et al. (\citeyear{Par08}),
The Cancer Genome Atlas (\citeyear{Net08})
and Jones et al. (\citeyear{Jon08,Jon10})] have sequenced the genomes of tumors in
order to find the causative or ``driver'' mutations. However, due to the
large number of genes being sequenced, one also finds a large number of
``passenger'' mutations that are genetically neutral and hence have no
role in the disease.

%
\begin{table}
\caption{Colorectal cancer data from Wood et al. (\citeyear{Woo07})}
\label{tab1}
\begin{tabular*}{\textwidth}{@{\extracolsep{\fill}}ld{3.0}d{3.0}ccc@{}}
\hline
& \multicolumn{2}{c}{\textbf{NS mutations}} & \multicolumn{3}{c@{}}{\textbf{Passenger probability}}
\\[-6pt]
& \multicolumn{2}{c}{\hrulefill} & \multicolumn{3}{c@{}}{\hrulefill} \\
\textbf{Gene} & \multicolumn{1}{c}{\textbf{Discovery}} & \multicolumn{1}{c}{\textbf{Validation}} & \textbf{External} & \textbf{SNP} & \textbf{NS/S} \\
\hline
APC & 171 & 138 & 0.00 &0.00 &0.00 \\
KRAS &79 &62 &0.00 &0.00 &0.00 \\
TP53 &79 &61 &0.00 &0.00 &0.00 \\
PIK3CA &28 &23 &0.00 &0.00 &0.00 \\
FBXW7 &14 &9 &0.00 &0.00 &0.00 \\
EPHA3 &10 &6 &0.00 &0.00 &0.00 \\
TCF7L2 &10 &7 &0.00 &0.00 &0.01 \\
ADAMTSL3  &9 &5 &0.00 &0.00 &0.03 \\
NAV3 &8 &3 &0.00 &0.01 &0.64 \\
GUCY1A2 &7 &4 &0.00 &0.00 &0.01 \\
MAP2K7 &6 &3 &0.00 &0.00 &0.02 \\[3pt]
PRKD1 &5 &3 &0.00 &0.00 &0.39 \\
MMP2 &5 &2 &0.00 &0.02 &0.61 \\
SEC8L1 &5 &2 &0.00 &0.03 &0.63 \\
GNAS &5 &2 &0.00 &0.04 &0.67 \\
ADAMTS18 &5 &2 &0.00 &0.07 &0.82 \\
RET &5 &2 &0.01 &0.17 &0.89 \\
TNN &5 &0 &0.00 &0.11 &0.81 \\
\hline
\end{tabular*}
\end{table}

To explain the issues involved in distinguishing the two types of
mutations, it is useful to take a look at a data set. Wood et
al. (\citeyear{Woo07}) did a ``discovery'' screen in which 18,191 genes were
sequenced in 11 colorectal cancers, and then a ``validation'' screen in
which the top candidates were sequenced in 96 additional tumors. The 18
genes that were mutated five or more times mutated in the discovery
screen are given in Table~\ref{tab1}. Here NS is short for nonsynonymous
mutation, a nucleotide substitution that changes the amino acid in the
corresponding protein. The top four genes in the list are well known to
be associated with cancer.

\begin{itemize}
\item
Adenomatous polyposis coli (APC) is a tumor suppressor gene. That is,
when both copies of the gene are knocked out in a cell, uncontrolled
growth results.
It is widely accepted that the first stages of colon cancer are the
loss of both copies of the APC gene
from some cell, see, e.g., Figure 4 in Luebeck and Moolgavkar (\citeyear{LueMol02}).
\item
Kras is an oncogene, i.e., one which causes trouble when a mutation
increases its expression level.
Once Kras is turned on it recruits and activates proteins necessary for
the propagation of growth factors.
\item
TP53 which produces the protein $p53$ (named for its 53 kiloDalton
size) is loved by those who study ``complex networks,'' since it is
known to be important and appears with very high degree in protein
interaction networks. $p53$ regulates the cell cycle and has been
called the ``master watchman'' referring to its role in conserving
stability by preventing genome mutation.
\item
The protein kinase PIK3CA is not as famous as the other three genes
(e.g., it does not yet have its own Wikipedia page)
but it is known to be associated with breast cancer. In a study of
eight ovarian cancer tumors in
Jones et al. (\citeyear{Jon10}), an $A \to G$ mutation was found at base 180,434,779 on
chromosome 3 in six tumors.
\end{itemize}

\noindent The next three genes on the list with the unromantic names FBXW7,
EPHA3, and TCF7L2 are all either known to be implicated in cancer or
are likely suspects because of the genetic pathways they are involved
in. Use google if you want to learn more about them.

The methodology that Wood et al. (\citeyear{Woo07}) used for assessing passenger
probabilities is explained in detail in
Parmigiani et al. (\citeyear{Par}). In principle this is straightforward: one
calculates the probability that the observed
number of mutations would be seen if all mutations were neutral. The
first problem is to
estimate the neutral mutation rate. In the column labeled ``external''
this estimate comes from
experimentally observed rates, while in the column labeled ``SNP'' they
used the mutations
observed in the study, with the genes declared to be under selection excluded.
The estimation problem is made more complicated by the fact that DNA
mutation rates are context dependent.
The two nucleotides in what geneticists call a CpG (the p refers to the
phosphodiester bond between the
adjacent cytosine and the guanine nucleotides) each mutate at roughly
10 times the rate of a thymine.

The third method for estimating passenger probabilities, inspired by
population genetics, is to look at
the ratio of nonsynonymous to synonymous mutations after these numbers
have been scaled by dividing by the number of opportunities for the two
types of mutations. While the top dozen genes show strong signals of
not being neutral,
as one moves down the list the situation becomes less clear, and the
probabilities reported in
the last three columns sometimes give conflicting messages.
The passenger probabilities in the last column are in most cases higher and
in some cases such as NAV3 and tthe last three genes in the table are
radically different. My personal feeling is that in this context the
NS/S test does not have enough mutations to give it power to detect
selection, but perhaps it is the other two methods that are being fooled.

To investigate the number and frequency of neutral mutations observed
in cancer sequencing studies, we will use a well-studied framework in
which an exponentially growing cancer cell population is modeled as a
multi-type branching process. Cells of type $i \ge0$ give birth at
rate $a_i$ and die at rate $b_i$, where the growth rate $\lambda
_i=a_i-b_i>0$. Thinking of cancer we will restrict our attention to the
case in which $i \to\lambda_i$ is increasing. To take care of
mutations, we suppose that individuals of type $i$ also give birth at
rate $u_{i+1}$ to individuals of type $i+1$ that have one more
mutation. This is slightly different from the approach of having
mutations with probability $u_{i+1}$ at birth, which translates into a
mutation rate of $a_i u_{i+1}$, and this must be kept in mind when
comparing with other results.

Let $\tau_k$ be the time of the first type $k$ mutation and let
$\sigma
_k$ be the time of the first type $k$
mutation that gives rise to a family that lives forever. Following up
on initial studies by Iwasa, Nowak and Michor (\citeyear{IwaNowMic06}), and Haeno, Iwasa
and Michor (\citeyear{HaeIwaMic07}), Durrett and Moseley (\citeyear{DurMos10}) have obtained results for
$\tau_k$ and limit theorems for the growth of $Z_k(t)$, the number of
type $k$'s at time~$t$. These authors did not consider $\sigma_k$, but
the extension is trivial: each type~$k$ mutation gives rise to a family
that lives forever with probability $\lambda_k/a_k$, so all we have to
do is to replace~$u_k$ in the limit theorem for $\tau_k$ by
$u_k\lambda_k/a_k$.

\subsection{Wave 0 results}\label{sec1.1}
To begin to understand the behavior of neutral mutations in our cancer
model, we first consider those that occur to type 0's, which are a
branching process $Z_0(t)$ in which individuals give birth at rate
$a_0$ and die at rate $b_0 < a_0$. It is well-known, see O'Connell
(\citeyear{OCo93}), that if we condition $Z_0(t)$ to not die out, and let $Y_0(t)$
be the number of individuals at time $t$ whose families do not die out,
then $Y_0(t)$ is a Yule process in which births occur at rate $\gamma
=\lambda_0/a_0$. Our first problem is to investigate the population
site frequency spectrum,
\begin{equation}
F(x) = \lim_{t\to\infty} F_t(x),
\label{Zsfsdef}
\end{equation}
where $F_t(x)$ is the expected number of neutral ``passenger''
mutations present in more than a fraction $x$ of the individuals at
time $t$. To begin to compute $F(x)$, we note that
\begin{equation}
Y_0(t)/Z_0(t) \to\gamma\qquad\mbox{in probability,}
\label{YoverZ}
\end{equation}
since each of the $Z_0(t)$ individuals at time $t$ has a probability
$\gamma$ of starting a family that does not die out, and the events are
independent for different individuals.

It follows from (\ref{YoverZ}) that it is enough to investigate the
frequencies of neutral mutations within $Y_0$. If we take the viewpoint
of the infinite alleles model, where each mutation is to a type not
seen before, then results can be obtained from Durrett and
Schweinsberg's (\citeyear{DurSch05}) study of a gene duplication model. In their
system there is initially a single individual of type 1. No individual
dies and each individual independently gives birth to a new individual
at rate 1. When a new individual is born it has the same type as its
parent with probability $1-r$ and with probability~$r$ is a new type
which is different from all previously observed types.

Let $T_N$ be the first time there are $N$ individuals and let $F_{S,N}$
be the number of families of size $>S$ at time $T_N$. Omitting the
precise error bounds given in Theorem 1.3 of Durrett and Schweinsberg
(\citeyear{DurSch05}), that result says
\begin{equation}
F_{S,N} \approx r \Gamma\biggl( \frac{2-r}{1-r} \biggr) N S^{-1/(1-r)}\qquad
\mbox{for $1 \ll S \ll N^{1-r}$}.
\label{DSF}
\end{equation}
The upper cutoff on $S$ is needed for the result to hold. When $S \gg
N^{1-r}$, $EF_{S,N}$ decays exponentially fast.

As mentioned above, the last conclusion gives a result for a branching
process with mutations according to the infinite alleles model, a
subject first investigated by Griffiths and Pakes\vadjust{\goodbreak} (\citeyear{GriPak88}). To study DNA
sequence data, we are more interested in the frequencies of individual
mutations. Using ideas from Durrett and Schweinsberg (\citeyear{DurSch04}) it is easy
to show:

\begin{theorem} \label{Z0sfs}
If passenger mutations occur at rate $\nu$ then $F(x) = \nu/\gamma x$.
\end{theorem}

This theorem describes the population site frequency spectrum. As in
Section~1.5 of Durrett (\citeyear{Dur}), this can be used to derive the site
frequency spectrum for a sample of size $n$. Let $\eta_{n,m}$ be the
number of sites in a sample of size $n$ where~$m$ individuals in the
sample have the mutant nucleotide. If one considers the Moran model in
a population of constant size $N$ then
\begin{equation}
E\eta_{n,m} = \frac{2N\nu}{m}\qquad \mbox{for $1 \le m < n$.}
\label{sfscsize}
\end{equation}
Using Theorem~\ref{Z0sfs} now, we get a new result concerning the
population genetics of exponentially growing populations. Here we are
considering a Moran model in an exponentially growing population, see,
e.g., Section 4.2 of Durrett (\citeyear{Dur}), rather than a branching process.

\begin{theorem} \label{expgrsfs}
Suppose that the mutation rate is $\nu$ and the population size $t$
units before the present is $N(t) = N e^{-\gamma t}$ then as $N\to
\infty$
\begin{equation}\label{Eetanm}
E\eta_{n,m}
\cases{\displaystyle \to \frac{n\nu}{\gamma} \cdot\frac
{1}{m(m-1)} , & \quad $2 \le m < n,$ \vspace*{2pt}\cr
\displaystyle\sim\frac{n\nu}{\gamma} \cdot\log(N\gamma),  &\quad$m=1,$}
\end{equation}
where $a_N\sim b_N$ means $a_N/b_N\to1$.
\end{theorem}

To explain the result for $m=1$, we note that, as Slatkin and Hudson
(\citeyear{SlaHud91}) observed, genealogies in exponentially growing
population tend to be star-shaped. The time required for $Y_0(t)$ to
reach size $N\gamma$ (and hence roughly the time for $Z_0(t)$ to reach
size $N$) is $\sim(1/\gamma) \log(N\gamma)$, so the number of
mutations on our $n$ lineages is roughly $n\nu$ times this. Note that,
(i) for a fixed sample size, $E\eta_{n,m}$, $2 \le m < n$ are bounded
independent of the final population size, and (ii) in contrast to (\ref
{sfscsize}), the sample size replaces the population size in formula
(\ref{Eetanm}).

The result in Theorem~\ref{expgrsfs} is considerably simpler than
previous formulas. Let $L(t)$ be the number of lineages $t$ units of
time before the present. For $2 \le k \le n$ let $T_k = \sup\{ t \dvtx  L(t)
\ge k\}$ be the first time at which the number of lineages is reduced
to $k-1$, and let $S_k=T_k-T_{k+1}$ where $T_{n+1}=0$. Griffiths and
Tavar\'e (\citeyear{GriTav98}) have shown that under some mild assumptions (coalescent
times have continuous distributions, only two lineages coalesce at
once, all coalescence events have equal probability, Poisson process of
mutations) the probability that a segregating site has $b$ mutant bases is
\begin{equation}\label{qnbGT}
q_{n,b} = \frac{(n-b-1)!(b-1)! \sum_{k=2}^n k (k-1) {{n-k}\choose{b-1}}
ES_k }{(n-1)! \sum_{k=2}^n k ES_k }.\vadjust{\goodbreak}
\end{equation}
To apply this result to the coalescent with population size $N(t) = N
e^{-\gamma t}$, one needs formulas for $ES_k$. See for example (52) in
Polanski, Bobrowski, and Kimmel (\citeyear{PolBobKim03}). However, these formulas are
complicated and difficult to evaluate numerically, since they involve
large terms of alternating size.
To connect (\ref{qnbGT}) with the result in Theorem~\ref{expgrsfs},
we write
\[
q_{n,1} = 1 - \frac{\sum_{k=2}^{n-1} k(n-k) ES_k}{(n-1) \sum_{k=2}^n
k ES_k}.
\]
Equation~(\ref{ESkO1}) below will show that $ES_n\sim\log N$ while for $2\le k
< n$, $ES_k = O(1)$ so
we have $1-q_{n,1} = O(1/\log N)$ in agreement with (\ref{Eetanm}).

\begin{figure}

\includegraphics{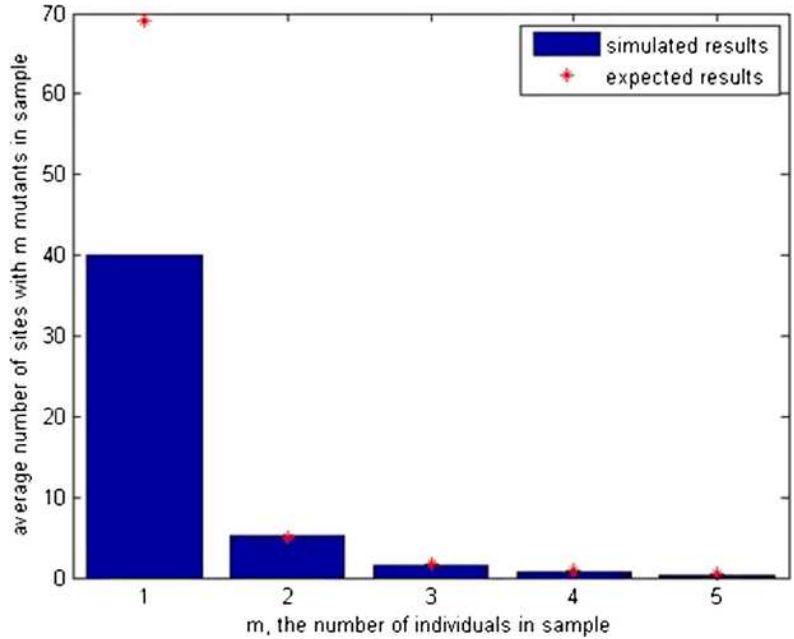}

\caption{Simulated site frequency spectrum when $\nu=\gamma$, sample
size $n=10$, and population size $N=100\mbox{,}000$.}\label{figsfs}
\end{figure}

To check (\ref{Eetanm}) Yifei Chen, a participant in a summer REU
associated with Duke's math biology
Research Training Grant, performed simulations. Figure~\ref{figsfs}
gives results for the average of 100 simulations with the indicated
parameters. The agreement is almost perfect for $m \ge2$ but the
formula considerably over
estimates the number of singletons with (\ref{Eetanm}), predicting
69.07 versus an observed value of
about~40. Given the approximations used in the proof of Theorem \ref
{expgrsfs} in Section~\ref{sec2}
for the case $m=1$, this is not surprising. The next result derives a
much better result for $E\eta_{n,1}$ which
gives a value of 36.66. See (\ref{YCans}) for details of the numerical
calculation.

\begin{theorem} \label{exactsfs}
\[
E \eta_{n,1} \approx\frac{\nu}{\gamma} \sum_{k=1}^{N\gamma}
\frac
{n}{n+k} \cdot\frac{k}{n+k-1}.
\]
\end{theorem}

Here $\approx$ means simply that this is an approximation which is
better for finite $N$.
As $N\to\infty$ the right-hand side $\sim(n\nu/\gamma) \log
(N\gamma)$
the answer in
Theorem~\ref{expgrsfs}.

The results for $E\eta_{n,m}$ are useful for population genetics, but
are not really relevant to cancer modeling. To investigate genetic
diversity in the exponentially growing population of humans, you would
sequence the DNA of a sample of individuals from the population.
However, in the study of cancer each patient has their own
exponentially growing cell population, so it is more interesting to
have the information provided by Theorem~\ref{Z0sfs} about the fraction
of cells in the population with a given mutation.

\begin{nm*} To illustrate the use of Theorem~\ref{Z0sfs}
suppose $\gamma= \lambda_0/a_0 =0.01$ and $\nu= 10^{-5}$. In support
of the numbers we note that Bozic et al. (\citeyear{Bozetal10}) estimate that the
selective advantage provided by a typical cancer driver mutation is
$0.004 \pm0.0004$. As for the second, if the per nucleotide mutation
rate is $10^{-8}$ and there are 1000 nucleotides in a gene then a
mutation rate of $10^{-5}$ per gene results. In this case Theorem \ref
{Z0sfs} predicts if we focus only on one gene then the expected number
of mutations with frequency $> 0.1$ is
\begin{equation}
F(0.1) = 10^{-5+2+1} = 0.01
\label{ce1}
\end{equation}
so, to a good first approximation, no particular neutral mutation
occurs with an appreciable frequency. Of course, if we are sequencing
20,000 genes then there will be a few hundred passenger mutations seen
in a given individual. On the other hand there will be very few
specific neutral mutations that will appear multiple times in the sample.
\end{nm*}
\subsection{Wave 1 results}\label{sec1.2}
We refer to the collection of type $k$ individuals as wave~$k$. In
order to analyze the cancer data, we also need results for neutral
mutations in waves $k>0$ of the multitype branching process. We begin
by recalling results from Durrett and Moseley (\citeyear{DurMos10}) for type 1
individuals in the process with $Z_0(0)=1$ when we condition the event
$\Omega^0_\infty$ that the type 0's do not die out. Let $\sigma_1$ be
the time of the first ``successful'' type 1 mutation that gives rise to
family that does not die out. Then $\sigma_1$ has median
\begin{equation}
s^1_{1/2} = \frac{1}{\lambda_0} \log\biggl( \frac{\lambda_0^2
a_1}{a_0u_1\lambda_1} \biggr)
\label{sig1med}
\end{equation}
and as $u_1 \to0$
\begin{equation}
P( \sigma_1 > s^1_{1/2} + x/\lambda_0 ) \to(1+e^x)^{-1}.
\label{sig1lim}\vadjust{\goodbreak}
\end{equation}
For (\ref{sig1med}) see (7) in Durrett and Moseley (\citeyear{DurMos10}) and drop the
$1$ inside the logarithm. The second result follows from the reasoning
for (6) there.

In investigating the growth of type 1's, it is convenient
mathematically to assume that $Z^*_0(t) = V_0 e^{\lambda_0 t}$
for $t \in(-\infty,\infty)$ and to let $Z_k^*(t)$ be the number of
type $k$'s at time $t$ in this system.
Here the star is to remind us that we have extended $Z_0$ to negative times.
The probability of a mutation to type 1 at times $t\le0$ is $\le V_0
u_1/\lambda_0$. In the concrete example
$u_1/\lambda_0=10^{-3}$, so this is likely to have no effect. The last
calculation omits two details that almost
cancel out. When we condition on survival of the type 0's,
$EV_0=a_0/\lambda_0$, but the probability a type 1 mutation
survives for all time is $\lambda_1/a_1$. Since $a_0 \approx a_1$ we
are too low by a factor of $\lambda_1/\lambda_0=2$.

Durrett and Moseley (\citeyear{DurMos10}) have shown:

\begin{theorem} \label{Z1lt}
If we regard $V_0$ as a fixed constant then as $t\to\infty$,
$e^{-\lambda_1 t}\times Z_1^*(t) \to V_1$ where $V_1$ is the sum of the points
in a Poisson process with mean measure $\mu(x,\infty) = c_{\mu,1} u_1
V_0 x^{-\alpha}$
with $\alpha= \lambda_0/\lambda_1$ and
\begin{equation}
c_{\mu,1} = \frac{1}{a_1} \biggl(\frac{a_1}{\lambda_1}\biggr)^{\sqz\alpha
} \Gamma(\alpha).
\label{cmu1def}
\end{equation}
The Laplace transform $E(e^{-\theta V_1}|V_0) = \exp( - c_{h,1} u_1 V_0
\theta^\alpha)$ where
$c_{h,1} =\break c_{\mu,1} \Gamma(1-\alpha)$. If $V_0$ is
$\operatorname{exponential}(\lambda
_0/a_0)$ then
\begin{equation}
E\exp(-\theta V_1) = \bigl(1+c_{h,1}u_1(a_0/\lambda_0)\theta^\alpha\bigr)^{-1}.
\label{LTV1}
\end{equation}
\end{theorem}

Here, and in what follows, constants like $c_{\mu,1}$, $c_{h,1}$, and
$c_{\theta,1}$ will depend on the branching process parameters $a_i$
and $b_i$, but not on the mutation rates $u_i$. The constant here is
equal to, but written differently from, the one in Durrett and Moseley
\[
c_{h,1} = \frac{1}{\lambda_0} \biggl(\frac{a_1}{\lambda_1}\biggr)^{\sqz
\alpha-1} \Gamma(1+\alpha)\Gamma(1-\alpha)
= \frac{1}{a_1} \frac{\lambda_1}{\lambda_0}\biggl (\frac{a_1}{\lambda
_1}\biggr)^{\sqz\alpha}
\alpha\Gamma(\alpha)\Gamma(1-\alpha).
\]
To prepare for later results note that the formula for the Laplace
transform shows that conditional on $V_0$, $V_1$ has a one sided stable
distribution with index $\alpha$.

The point process in Theorem~\ref{Z1lt} describes the contributions of
the successful type 1 mutations to $Z_1(t)$. The first such mutation
occurs at time $\sigma_1$, which has median $s^1_{1/2}$. The derivation
of Theorem~\ref{Z1lt} is based on the observation that a mutation at
time $s$ will grow to size $\approx e^{\lambda_1(t-s)} W_1$ by time~$t$,
where $W_1$ has distribution
\[
W_1 =_d \frac{b_1}{a_1} \delta_0 + \frac{\lambda_1}{a_1} \operatorname{exponential}(\lambda_1/a_1)
\]
and hence make a contribution of $e^{-\lambda_1 (s - s^1_{1/2})}$ to
the limit $\bar V_1$.
Thus we expect that most of the mutations that make a significant
contribution will come within a time $O(1/\lambda_1)$ of $s^1_{1/2}$.\vadjust{\goodbreak}

The complicated constants in Theorem~\ref{Z1lt} can be simplified if we
instead look at the limit
\[
e^{-\lambda_1(t-s^1_{1/2})} Z_1^*(t) \to\bar V_1 =_d V_1 \exp
(\lambda
_1 s^1_{1/2}).
\]
Using the definition of $s^1_{1/2}$ in (\ref{sig1med}) and recalling
$\alpha=\lambda_0/\lambda_1$ we see that
\[
\exp(\lambda_1 s^1_{1/2}) = \biggl( \frac{\lambda_0 a_1}{a_0 u_1} \cdot
\alpha\biggr)^{\sqz1/\alpha}
\]
and hence using (\ref{LTV1})
\begin{equation}
E\exp(-\theta\bar V_1)
= \biggl( 1 + \alpha\Gamma(\alpha) \Gamma(1-\alpha) \biggl(\frac{a_1
\theta}{\lambda_1}\biggr)^{\sqz\alpha} \biggr)^{-1}.
\label{LTbarV1}
\end{equation}
The combination of Gamma functions is easy to evaluate, since Euler's
reflection function
implies that
\begin{equation}
\alpha\Gamma(\alpha) \Gamma(1-\alpha) = \frac{\pi\alpha}{\sin
(\pi\alpha
)} > 1.
\label{erf}
\end{equation}
A second look at (\ref{LTbarV1}) shows that $a_1\bar V_1/\lambda_1$ has
a distribution that only depends on
$\alpha$. For comparison, note that if $V_0$ is $\operatorname{exponential}(\lambda
_0/a_0)$ then $a_0V_0/\lambda_0$ is
$\operatorname{exponential}(1)$.

Using results for one-sided stable laws, Durrett et al. (\citeyear{Duretal11}) were able to prove results about the genetic
diversity of wave 1. Define Simpson's index to be the limiting
probability two randomly chosen individuals in wave 1 are descended
from the same type 1 mutation. In symbols, it is the $p=2$ case of the
following definition
\[
R_p = \sum_{i=1}^\infty\frac{X_i^p}{V_1^p},
\]
where $X_1 > X_2 > \cdots$ are points in the Poisson process and $V_1$
is the sum.
The result for the mean, which comes from a result of Fuchs, Joffe and
Teugels (\citeyear{FJT01}), is much simpler than one could reasonably expect.

\begin{theorem} \label{ESI}
$ER_2 = 1 - \alpha$.
\end{theorem}

After this paper was written Jason Schweinsberg explained to me that
the points $Y_i = X_i/V_1$ have the Poisson--Dirichlet distribution
$\mathrm{PD}(\alpha,0)$, so Theorem~\ref{ESI} follows from (3.6) in Pitman
(\citeyear{Pit06}). For our purposes it is easier to refer to (6) in Pitman and Yor
(\citeyear{PitYor97}) where it is shown that
\[
E \sum_{i=1}^\infty f(Y_i) = \frac{1}{\Gamma(\alpha) \Gamma
(1-\alpha)}
\int_0^1 f(u) u^{-\alpha-1} (1-u)^{\alpha-1}.
\]
Taking $f(x)=x^p$ we find that $R_p = \sum_i X_i^p/V_k^p$ has
\[
ER_p = E \sum_i Y_i^p = \frac{\Gamma(p-\alpha)}{\Gamma(1-\alpha)
\Gamma(p)}.
\]

Using formulas in Logan et al. (\citeyear{Lo73}) one can derive
results for the distribution of $R_2^{-1/2}$. Work of
Darling (\citeyear{Da52}) leads to information about the distribution of the
fraction in the largest clone $X_1/V_1$.
In particular,

\begin{theorem} \label{maxclone}
$V_1/X_1$ has mean $1/(1-\alpha)$.
\end{theorem}

Since $1/x$ is convex, $E(X_1/V_1) > 1/E(V_1/X_1) = 1-\alpha$.

Theorems~\ref{ESI} and~\ref{maxclone} suggest that if we are interested
in understanding neutral mutations in say 90\% of the population when
wave 1 is dominant, then we can restrict our attention to the families
generated by a small number of the most prolific type~1 mutants. (The
number we need to consider will be large if $\alpha$ is close to 1.)
The result in (\ref{ce1}) suggests that we can ignore neutral mutations
within the descendants of these type~1 mutations. Mutations that occur
on the genealogies of the $i$th largest mutations will appear in all of
their descendants and hence have frequency $X_i/V_1$. As remarked above
(and explained in more detail in Section~\ref{sec3}), the genealogies of the
most prolific type 1 mutants will be approximately star-like so they
will mostly have different mutations. Note that here, in contrast to
the reasoning that led to (\ref{err1}) there are several individuals
founding different subpopulations whose genealogies have collected
neutral mutations.

\subsection{Wave $k$ results}\label{sec1.3}

Once Theorem~\ref{Z1lt} was established it was straightforward to
extend the result by induction.
Let $\alpha_k=\lambda_{k-1}/\lambda_k$,
\begin{equation}
c_{\mu,k} = \frac{1}{a_{k}} \biggl( \frac{a_k}{\lambda_k} \biggr)^{\alpha
_k} \Gamma(\alpha_k)\quad
\mbox{and}\quad c_{h,k} = c_{\mu,k} \Gamma(1-\alpha_k).
\label{chkdef}
\end{equation}
Let $c_{\theta,0} = a_0/\lambda_0$, $\mu_0=1$ and inductively define
for $k \ge1$
\begin{eqnarray}
c_{\theta,k} & = &c_{\theta,k-1} c_{h,k}^{\lambda_0/\lambda_{k-1}},
\label{cthrec}
\\
\mu_k & =& \mu_{k-1} u_k^{\lambda_0/\lambda_{k-1}} = \prod_{j=1}^k
u_j^{\lambda_0/\lambda_{j-1}}.
\label{mukrec}
\end{eqnarray}

Durrett and Moseley (\citeyear{DurMos10}) have shown:

\begin{theorem} \label{Zklimth}
Suppose $Z^*_0(t) = V_0 e^{\lambda_0t}$ for $t\in(-\infty,\infty)$
where $V_0$ is\break $\operatorname{exponential}(\lambda_0/a_0)$.
\[
e^{-\lambda_k t} Z^*_k(t) \to V_k \qquad\mbox{a.s.}
\]
Let $\mathcal{ F}^{k-1}_\infty$ be the $\sigma$-field generated by
$Z^*_j(t)$, $j \le k-1$, $t\ge0$. $(V_k|\mathcal{ F}^{k-1}_\infty)$
is the sum of the points in a Poisson\vadjust{\goodbreak} process with mean measure $\mu
(x,\infty) = c_{\mu,k} u_k V_{k-1} x^{-\alpha_k}$.
\[
E(e^{-\theta V_k}|\mathcal{ F}^{k-1}_\infty) = \exp(-c_{h,k} u_k V_{k-1}
\theta^{\lambda_{k-1}/\lambda_k})
\]
and hence
\begin{equation}
Ee^{-\theta V_k} = ( 1 + c_{\theta,k} \mu_k \theta^{\lambda
_0/\lambda_{k}} )^{-1}.
\label{ltVk}
\end{equation}
\end{theorem}

Using Theorem~\ref{Zklimth} it is easy to analyze $\tau_{k+1}$, the
waiting time for the first type $k+1$.
Details of the derivations of (\ref{tmedeq}) and (\ref{skrecc}) are
given in Section~\ref{sec4}. The median of $\tau_{k+1}$ is
\begin{equation}
t^{k+1}_{1/2} = \frac{1}{\lambda_0} \log\biggl(
\frac{ \lambda_k^{\lambda_0/\lambda_k} }{ c_{\theta,k} \mu_{k+1} }
\biggr)
= \frac{1}{\lambda_k} \log(\lambda_k)
- \frac{1}{\lambda_0} \log( c_{\theta,k} \mu_{k+1} )
\label{tmedeq}
\end{equation}
and as in the case of $\tau_1$
\[
P( \tau_{k+1} > t^{k+1}_{1/2} + x/\lambda_0 ) \approx( 1 + e^x )^{-1}.
\]
Again the result for the median $s^{k+1}_{1/2}$ of the time $\sigma
_{k+1}$ of the first mutation to type $k+1$ with a family that does not
die out can be found by replacing $u_{k+1}$ by $u_{k+1} \lambda_{k+1}/a_{k+1}$.

Formula (\ref{tmedeq}), due to Durrett and Moseley (\citeyear{DurMos10}), is not very
transparent due to the complicated constants. We will obtain a more
intuitive result by looking at the difference $s^{k+1}_{1/2} - s^k_{1/2}$.
After some algebra, hidden away in Section~\ref{sec4}, we have
\begin{equation}\qquad
s^{k+1}_{1/2} - s^k_{1/2} = \frac{1}{\lambda_k} \log\biggl( \frac{\lambda
^2_k a_{k+1} }{ a_k u_{k+1}\lambda_{k+1} } \biggr)
- \frac{1}{\lambda_{k-1}} \log\bigl( \alpha_k \Gamma(\alpha_k)\Gamma
(1-\alpha
_k) \bigr).
\label{skrecc}
\end{equation}

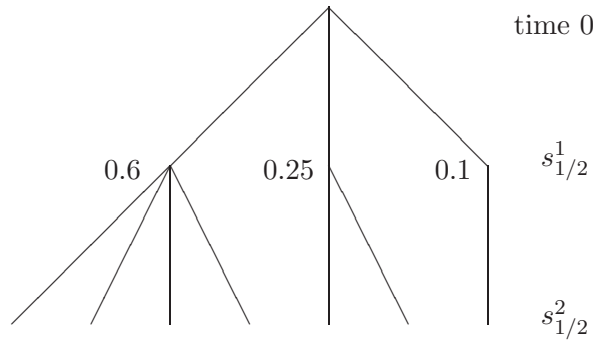
\begin{figure}
\begin{center}
\begin{picture}(310,150)
\put(140,140){\line(-1,-1){60}}
\put(140,140){\line(0,-1){60}}
\put(140,140){\line(1,-1){60}}
\put(80,80){\line(-1,-1){60}}
\put(80,80){\line(-1,-2){30}}
\put(80,80){\line(0,-1){60}}
\put(80,80){\line(1,-2){30}}
\put(140,80){\line(0,-1){60}}
\put(140,80){\line(1,-2){30}}
\put(200,80){\line(0,-1){60}}
\put(220,20){$s^2_{1/2}$}
\put(220,80){$s^1_{1/2}$}
\put(210,130){time 0}
\put(55,75){0.6}
\put(115,75){0.25}
\put(180,75){0.1}
\end{picture}
\caption{Genealogy of wave 2 individuals. Here 0.6, 0.25, and 0.1 are
the fractions of the
type 1 population derived from the three most prolific type 1
mutations. If these numbers look odd
recall that in the example $ER=1/2$ for wave 1, while $(0.6)^2 + (0.25)^2
+ (0.1)^2=0.4325$.}\label{fig2}
\end{center}
\end{figure}

\textit{Neutral mutations}. Returning to our main topic, it follows from
the first conclusion in
Theorem~\ref{Zklimth} that the results of Theorems~\ref{ESI} and \ref
{maxclone} hold for wave $k$
when~$\alpha$ is replaced by $\alpha_k = \lambda_{k-1}/\lambda_k$.
Suppose for simplicity that $k=2$.
In the concrete example $\alpha_2=2/3$, so $ER_2=1/3$ and again there
will be a small number
of type 2 mutations that occur at times close to $s^2_{1/2}$ that are
responsible for 90\% of the population. If we let $x_1 > x_2 > \cdots$
be the fractions of the
type 1 population that result from the most prolific type 1 mutants,
then the $j$th most prolific
type 2 mutation will trace its lineage back to the $i$th most prolific
type 1 mutation with
probability~$x_i$. All of the type 2 mutants who trace their ancestry
back to the same type~1
mutant will have lineages that coalesce at times near $s^1_{1/2}$.
Working backwards from that
time the genealogy of the most prolific type 1 mutations will be star
like. At this point a picture is worth a hundred words, see Figure~\ref{fig2}.

%

\subsection{\texorpdfstring{Relationship to Bozic et al. (\citeyear{Bozetal10})}{Relationship to Bozic et al. (2010)}}\label{sec1.4}
The inspiration for this investigation came from a paper by Bozic et
al. (\citeyear{Bozetal10}). Their model takes place in discrete time to facilitate
simulation and their types are numbered starting from 1 rather than\vadjust{\goodbreak}
from 0. At each time step, a cell of type $j\ge1$ either divides into
two cells, which occurs with probability $b_j$, or dies with
probability $d_j$ where $d_j=(1-s)^j/2$ and $b_j=1-d_j$. It is
unfortunate that their birth probability $b_j$ is our death rate for
type $j$ cells. We will not resolve this conflict because but we want
to preserve their notation make it easy to compare with the results in
the paper.

In addition, at every division, the new daughter cell can acquire an
additional driver mutation with probability $u$,
or a passenger mutation with probability $\nu$. They find the following
result for the expectation of $M_k$, the number of passenger mutations
in a tumor that has accumulated $k$ driver mutations:
\begin{equation}
EM_k = \frac{\nu}{2s} \log\frac{4ks^2}{u^2} \log k.
\label{Bozicf}
\end{equation}
The derivation of this formula suffers from two errors due to a
fundamental misconception, and loses accuracy because of some dubious
arithmetic.
The first error is to claim that (see Section 5 of their supplementary
materials)
\begin{equation}
EM_k = \frac{\nu}{T} E\sigma_k,
\label{err1}
\end{equation}
where $T$ is the average time between cell divisions. In essence (\ref
{err1}) asserts that the passenger mutations in the population are
exactly those that have appeared along the genealogy of the cell with
the first type $k$ mutation that gives rise to a family that lives
forever. However as Theorems~\ref{Z1lt} and~\ref{Zklimth} show, this is
wrong because after the initial wave more than one mutation makes a
significant contribution to the size of the type $k$ population.

The second erroneous ingredient is (S5) in their supplementary
materials. In quoting that result below we have dropped the $1+$ inside
the $\log$ in their formula, since it disappears in their later
calculations and this makes their result easier to relate to ours.
\begin{equation}
\qquad E(\sigma_{j+1} - \sigma_j) = \frac{T \log[ {(1-q_j)}/{(u
b_j(1-q_{j+1}))} ( 1 - {1}/{(b_j(2-u))} ) ]}
{\log[b_j(2-u)]},
\label{err2}
\end{equation}
where $q_j$ is probability that a type $j$ mutation dies out. By
considering what happens on the first step:
\begin{equation}
q_j \approx d_j + b_j q_j^2 \quad\mbox{and hence} \quad q_j \approx\frac
{d_j}{b_j} \approx\frac{1-js}{1+js} \approx1-2js,
\label{extpr}
\end{equation}
where the last approximation assumes that $s$ is small.

Before we start to compare results, recall that Bozic et al. (\citeyear{Bozetal10})
number their waves starting with 1 while our numbers start at 0. When
the differences in notation are taken into account (\ref{sig1med})
agrees with the $j=1$ case of (\ref{err2}). The death and birth
probabilities in the model of Bozic et al. (\citeyear{Bozetal10}) are $d_1=(1-s)/2$ and
$b_1=1-d_1=(1+s)/2$, so $\log(2b_1) \approx\log(1+s) \approx s$. $q_j
\approx(1-js)/(1+js) \approx1-2js$. Taking into account the fact that
mutations occur only in the new daughter cell at birth, we have
$u_1=b_1u$, so when $j=1$ (\ref{err2}) becomes
\[
E(\sigma_2-\sigma_1) \approx\frac{1}{s} \log\biggl( \frac{s^2}{u_1
\cdot2s }\biggr).
\]
Setting $\lambda_j = (j+1)s$, and $a_i=b_{i+1}$ in our continuous time
branching process, we have $a_1/a_0\approx1$ and this agrees with
(\ref{sig1med}).

\begin{nm*} To match a choice of parameters studied in
Bozic et al. (\citeyear{Bozetal10}), we will take $u=10^{-5}$ and $s=0.01$, so $u_i=b_i
u \approx5 \times10^{-6}$, and
\[
s^1_{1/2} \approx\frac{1}{0.01} \log\biggl( \frac{10^{-4}}{5 \times
10^{-6} \cdot0.02} \biggr)
= 100 \log(1000) = 690.77.
\]
Note that by (\ref{sig1lim}) the fluctuations in $\sigma_1$ are of
order $1/\lambda_0 = 100$.

To connect with reality, we note that for colon cancer the average time
between cell divisions is $T=4$ days,
so 690.77 translates into 7.57 years. In contrast, Bozic et al. (\citeyear{Bozetal10})
compute a waiting time of 8.3 years on page 18,546.
This difference is due to the fact that the formula they use [(1) on
the cited page] employs the approximation $1/2 \approx1$.

Turning to the later waves, we note that:

\begin{longlist}[(ii)]
\item[(i)] the first ``main'' term in (\ref{skrecc}) corresponds to the answer
in (\ref{err2}).

\item[(ii)] by (\ref{erf}), $\alpha_k \Gamma(\alpha_k)\Gamma(1-\alpha_k)
= \pi
\alpha_k/ \sin(\pi\alpha_k) > 1$, so the ``correction'' term not
present in (\ref{err2}) is $<0$, which is consistent with the fact that
the heuristic leading to (\ref{err2}) considers only the first
successful mutation.
\end{longlist}

To obtain some insight into the relative sizes of the ``main'' and the
``correction'' terms in (\ref{skrecc}), we will consider our concrete
example in which $\lambda_i = (i+1)s$ and $a_i=b_{i+1} \approx1/2$, so
for $i\ge1$
\[
s^{i+1}_{1/2}-s^i_{1/2} = \frac{1}{(i+1)s} \log\biggl(\frac
{(i+1)^2s}{u_{i+1}(i+2)}\biggr)
- \frac{1}{is} \log\biggl( \frac{\pi\alpha_i}{\sin(\pi\alpha_i)}
\biggr).
\]
Taking $s=0.01$, $u=10^{-5}$ and $u_i=5 \times10^{-6}$ leads to the
results given in Table~\ref{tab2}.

%
\begin{table}
\caption{Comparison of expected waiting times from (\protect\ref{skrecc}) and
(\protect\ref{err2}). The numbers in parentheses are the answers converted into
years using $T=4$ as the average number of days between cell divisions}\label{tab2}
\begin{tabular*}{\textwidth}{@{\extracolsep{\fill}}lcclcc@{}}
\hline
& \textbf{Main} & \textbf{Corr.} & & \textbf{From (\ref{skrecc})} & \textbf{From (\ref{err2})} \\
\hline
$s^1_{1/2}$ & 690.77 & 0\phantom{0.} & $s^1_{1/2}$ & 690.77 (7.57) & 550.87 (6.04)
\\[3pt]
$s^2_{1/2}-s^1_{1/2}$ & 394.41 & 45.15 & $s^2_{1/2}$ & 1040.03 (11.39)
& 895.39 (9.81) \\[3pt]
$s^3_{1/2}-s^2_{1/2}$ & 280.36 & 44.15 & $s^3_{1/2}$ & 1276.24 (13.98)
& 1149.79 (12.60)\\
\hline
\end{tabular*}
\end{table}

The values in the last column differ from the sum of the values in the
first column because Bozic et al. (\citeyear{Bozetal10})
indulge in some dubious arithmetic to go from their formula
\[
E(\sigma_{j+1} - \sigma_j) = \frac{1}{js} \log\biggl( \frac
{2j^2s}{(j+1)u} \biggr)
\]
to their final result
\[
E\sigma_k \approx\frac{1}{2s} \log\biggl( \frac{4ks^2}{u^2} \biggr)
\log k.
\]
First they use the approximation $j/(j+1) \approx1$ and then $\sum
_{j=1}^{k-1} \approx\int_0^k$.
In the first row of the table this means that their formula
underestimates the right answer by 20\%.
Bozic et al. (\citeyear{Bozetal10}) tout the excellent agreement between their formula
and simulations given in their Figure S2. However, a closer look at the
graph reveals that while their formula underestimates simulation
results, our answers agree with them almost exactly.
\end{nm*}

\section{Proofs for wave 0}\label{sec2}
\mbox{}
\begin{pf*}{Proof of Theorem \protect\ref{Z0sfs}}
Dropping the subscript 0 for convenience, recall that $Y(t)$ is defined
to be the number of individuals in the branching
process $Z(t)$ with an infinite line of descent and that $Y(t)$ is a
Yule process with birth rate $\gamma=\lambda_0/a_0$. For $j\ge1$ let
$T_j = \min\{ t \dvtx  Y_t = j \}$ and notice that $T_1=0$. Since the $j$
individuals at time $T_j$ start independent copies $Y^1, \ldots, Y^j$ of
$Y$, well known results for the Yule process imply
\[
\lim_{s\to\infty} e^{-\gamma s} Y^i(s) = \xi_i,
\]
where the $\xi_i$ are independent exponential mean 1 (here time $s$ in
$Y^i$ corresponds to time $T_j+s$ in the original process). From the
limit theorem for the $Y^i$ we see that for $j\ge2$ the limiting
fraction of the population descended from individual $i$ at time $T_j$ is
\[
r_i = \xi_i/(\xi_1 + \cdots+ \xi_j),\qquad 1 \le i \le j
\]
which as some of you know has a beta$(1,j-1)$ distribution with density
$(j-1)(1-x)^{j-2}$.

To prepare for the simulation algorithm it is useful to give an
explicit proof of this fact. Note that
\[
\bigl((\xi_1, \ldots,\xi_j)|\xi_1 + \cdots+ \xi_j=t\bigr)
\]
is uniform over all nonnegative vectors that sum to $t$, so
$(r_1, \ldots, r_j)$ is uniformly distributed over the nonnegative
vectors that sum to 1. Now the joint distribution of the $r_i$ can be
generated by letting $U_1, \ldots ,U_{j-1}$ be uniform on $[0,1]$,
$U^{(1)} < U^{(2)} < \cdots< U^{(j-1)}$ be the order statistics, and
$r_i = U^{(i)}-U^{(i-1)}$ where $U^{(0)}=0$ and $U^{(j)}=1$. From this
and symmetry, we see that
\[
P( r_i > x) = P(r_j>x) = P( U_i < x \mbox{ for $1\le i \le j-1$}) = (1-x)^{j-1}
\]
and differentiating gives the density.

If the neutral mutation rate is $\nu$ then on $[T_j,T_{j+1})$ mutations
occur to individuals in $Y$ at rate $\nu j$, while births occur at rate
$\gamma j$, so the number of mutations $N_j$ in this time interval has
a shifted geometric distribution with success probability $\gamma
/(\gamma+\nu)$, i.e.,
\begin{equation}
P( N_j = k) = \biggl( \frac{\nu}{\nu+\gamma} \biggr)^k \frac{\gamma}{\nu
+\gamma}\qquad
\mbox{for $k=0,1,2, \ldots.$}
\label{shgeom}
\end{equation}
The $N_j$ are i.i.d. with mean
\[
\frac{\nu+\gamma}{\gamma} -1 = \frac{\nu}{\gamma}.
\]
Thus the expected number of neutral mutations that are present at
frequency larger than $x$ is
\[
\frac{\nu}{\gamma} \sum_{j=1}^\infty(1-x)^{j-1} = \frac{\nu
}{\gamma x}.
\]
The $j=1$ term corresponds to mutations in $[T_1,T_2)$ which will be
present in the entire population.
\end{pf*}

\textit{Simulation algorithm.} The proof of the last result leads to a
useful simulation algorithm.
Suppose we have worked our way up to time $T_j$ with $j \ge1$ and the
limiting fractions of
the descendants of the $j$ individuals at this time correspond to the
sizes of the intervals
\[
0 = U_{j,0} < U_{j,1} < \cdots< U_{j,j-1} < U_{j,j} = 1,
\]
where the $U_{j,i}$, $1\le i<j$, are the order statistics of a sample
of $j-1$ independent uniforms.

To take care of mutations in $[T_j,T_{j+1})$, we generate a number of
mutations $N_j$ with a shifted geometric distribution given in (\ref
{shgeom}) and associate each mutations with an interval $(U_{j,i-1},
U_{j,i})$ with $i$ chosen at random from $1, \ldots, j$.

To produce the subdivision at time $T_{j+1}$, let $V$ be an independent
uniform, define $1\le n_j\le j$
so that $U_{j,n_j-1} < V < U_{j,n_j}$, and then let
\[
U_{j+1,i} =
\cases{ U_{j,i}, & \quad $0 \le i < n_j,$ \vspace*{2pt}\cr
V, & \quad $i = n_j,$ \vspace*{2pt}\cr
U_{j,i-1} ,& \quad $n_j
< i \le j+1$.}
\]
Note that the interval to be split is not chosen at random but
according to its length.
The simplest explanation of why this is true is that it is needed to
have the new point added
be uniform on $(0,1)$. For a detailed explanation, see Theorem 1.8 of
Durrett (\citeyear{Dur}).

When we have worked our way down to $T_j$ with $j=N\gamma$ we stop. To
find the properites of a sample of
size $n$, we choose points $X_1, \ldots, X_n$ independently and uniform
on $(0,1)$. For each $k$ a mutation
associated with $(U_{k,i-1},U_{k,i})$ appears in all of the individual
$X_m \in(U_{k,i-1},U_{k,i})$.

\begin{pf*}{Proof of Theorem \protect\ref{expgrsfs}}
We begin with a calculus fact, that is, easy for readers who can remember
the definition of the beta distribution. The rest of us can simply
integrate by parts.

\begin{lemma} If $a$ and $b$ are nonnegative integers
\begin{equation}
\int_0^1 x^a (1-x)^b \,dx = \frac{a! b!}{(a+b+1)!}.
\end{equation}
\end{lemma}

Differentiating the distribution function from Theorem~\ref{Z0sfs}
gives the density $\nu/\gamma x^2$. We have removed the atom at 1 since
those mutations will be present in every individual and we are
supposing the sample size $n > m$ the number of times the mutation
occurs in the sample.
Conditioning on the frequency in the entire population, it follows that
for $m\le2 < n$ that
\[
E\eta_{n,m} = \int_0^1 \frac{\nu}{\gamma x^2} \pmatrix{{n}\vspace*{2pt}\cr{m}} x^m
(1-x)^{n-m} \,dx
= \frac{n\nu}{\gamma m(m-1)},
\]
where we have used $n\ll N$ and the second step requires $m \ge2$.\vadjust{\goodbreak}

When $m=1$ the formula above gives $E\eta_{n,1}=\infty$. To get a
finite answer we note that
$Z_t=n$ roughly when $Y_t =n\gamma$ so the expected number that are
present at frequency larger than $x$ is
\[
\frac{\nu}{\gamma} \sum_{j=1}^{N\gamma} (1-x)^{j-1} = \frac{\nu
}{\gamma
x}\bigl ( 1 - (1-x)^{N\gamma} \bigr).
\]
Differentiating (and multiplying by $-1$) changes the density from $\nu
/\gamma x^2$ to
\begin{equation}
\frac{\nu}{\gamma} \biggl( \frac{1}{x^2} \bigl( 1 - (1-x)^{N\gamma}
\bigr)
- \frac{1}{x} N\gamma(1-x)^{N\gamma-1} \biggr).
\label{trdens}
\end{equation}
Ignoring the constant $\nu/\gamma$ for the moment and noticing
${{n}\choose{m}} x^m (1-x)^{n-m} = nx(1-x)^{n-1}$
when $m=1$ the contribution from the second term is
\[
n \int_0^1 N\gamma(1-x)^{N\gamma+n-2} \,dx = n \cdot\frac{N\gamma
}{N\gamma+n-1} < n
\]
and this term can be ignored. Changing variables $x=y/N\gamma$ the
first integral is
\begin{eqnarray*}
&&\int_0^1 \frac{1}{x} \bigl(1-(1-x)^{N\gamma}\bigr) (1-x)^{n-1} \,dx
\\
&&\qquad= \int_0^{N\gamma} \frac{1}{y} \bigl(1-(1-y/N\gamma)^{N\gamma}\bigr)
(1-y/N\gamma)^{n-1} \,dy.
\end{eqnarray*}
To show that the above is $\sim\log(N\gamma)$ we let $K_N\to\infty$
slowly and divide the integral into three regions
$[0,K_N]$, $[K_N,N\gamma/\log N]$, and $[N\gamma/\log N,N\gamma]$.
Oustide the first interval,
$(1-y/N\gamma)^{N\gamma} \to0$ and outside the third, $(1-y/N\gamma
)^{n-1} \to1$ so we conclude that the above is
\[
O(K_N) + \int_{K_N}^{N\gamma/\log N} \frac{1}{y} \,dy + O(\log\log N).
\]
As the simulation results cited in the introduction suggest, this
approximation is somewhat rough.
\end{pf*}

\begin{pf*}{Proof of Theorem \protect\ref{exactsfs}}
When a mutation that occurs on level $j=k+1$ is associated with
$(U_{j,i-1},U_{j,i})$ it affects
all members of the sample that land in that interval. By symmetry of
the joint distribution of
the interval lengths, we can suppose without loss of generality that
$i=1$. Think of the
$k$ break points $U_{j,i}$ with $1<i<j-1$ as red points and the $n$
uniforms $X_1, \ldots ,X_n$
as blue. The mutation will affect exactly one individual in the sample
if as we look
from left to right, the first point is blue and the second is red. By
symmetry this
has probability
\[
\frac{n}{n+k} \cdot\frac{k}{n-1+k}.
\]
Taking into account that the mean number of mutations per level is $\nu
/\gamma$ and summing gives
desired formula.
\end{pf*}

\textit{Evaluating the constant}. Writing $M$ for $N\gamma$,
\begin{eqnarray*}
\sum_{k=1}^M \frac{n}{n+k} \cdot\frac{k}{n-1+k} &=& n \sum_{k=1}^M
\frac{1}{n+k}
\cdot\biggl( 1- \frac{n-1}{n-1+k} \biggr) \\
& =& n \sum_{j=n+1}^{n+M} \frac{1}{j} - n(n-1) \sum_{k=1}^M \biggl( \frac
{1}{n+k-1} - \frac{1}{n+k} \biggr).
\end{eqnarray*}
The second sum telescopes and has value
\[
- n(n-1) \biggl( \frac{1}{n} - \frac{1}{n+M} \biggr) \approx-(n-1).
\]
If $\rho$ is Euler's constant then the first sum is
\[
\approx\log(n+M) + \rho- \sum_{j=1}^n \frac{1}{j}.
\]
If $n=10$ and $M=1000$ then we end up with
\begin{equation}
10 \cdot[6.9177 + 0.5772 - 2.929] - 9 = 36.66.
\label{YCans}
\end{equation}

\section{Genealogies}\label{sec3}
A simple description and a useful mental picture of genealogies in an
exponentially
growing population is provided by the following result of Kingman (\citeyear{Kin}).

\begin{theorem} \label{timech}
If we run time at rate $1/N(s)$ then on the new time scale genealogies
follow the standard coalescent in which
there is coalescence at rate ${{k}\choose{2}}$ when there are $k$ lineages.
\end{theorem}

When $N(t)=N e^{-\gamma t}$ the time interval $[0,(1/\gamma)\log N)$
over which the model makes sense gets mapped by the time change to an
interval of length
\[
\frac{1}{N} \int_0^{(1/\gamma)\log N} e^{\gamma t} \,dt = \frac
{1}{\gamma} \cdot\frac{N-1}{N} < \frac{1}{\gamma}.
\]

While Theorem~\ref{timech} is useful conceptually, it is difficult to
use for computations because after the
time change mutations occur at a time-dependent rate. Back on the
original time scale, Griffiths and
Tavar\'e (\citeyear{GriTav98}) have shown that the joint density of the coalescent
times $(T_k, \ldots, T_n)$
for any $k\ge2$ is given by
\begin{equation}
p_{k,n}(t_k, \ldots, t_n) = \prod_{j=k}^n \frac{ {{j}\choose{2}} }{
N(t_j) }
\exp\biggl( - \int_{ t_{j+1} }^{t_j} \frac{ {{j}\choose{2}} }{ N(s) } \,ds
\biggr),
\label{jtdistco}
\end{equation}
where $0 = t_{n+1} < t_n <\cdots< t_k$. In particular when $k=n$ and
$N(t)=N e^{-\gamma t}$
\begin{equation}
p_n(t_n) = \frac{n(n-1)}{2N} e^{\gamma t_n} \exp\biggl( - \frac
{n(n-1)}{2N\gamma} (e^{\gamma t_n} - 1) \biggr).
\label{firstco}
\end{equation}

One can, in principle at least, find the marginal distribution $p_k$ of
$t_k$ by integrating out the
variables $t_{k+1}, \ldots, t_n$ in (\ref{jtdistco}). According to
(5)--(8) in Polanski, Bobrowski, and Kimmel (\citeyear{PolBobKim03})
\begin{eqnarray}
\label{PBK}
p_k(t_k)  = \sum_{j=k}^n A^k_j q_j(t_k)\qquad \mbox{where}
\nonumber
\\[-8pt]
\\[-8pt]
\eqntext{q_j(t_k)  = \displaystyle\frac{ {{j}\choose{2}} }{ N(t_k) }
\exp\biggl( - \int_{0}^{t_k} \frac{ {{j}\choose{2}} }{ N(s) } \,ds
\biggr)}
\end{eqnarray}
and the coefficients $A^k_j$ are given by $A^n_n=1$
\[
A^k_j = \frac{ \prod_{\ell=k, \ell\neq j}^n {{\ell}\choose{2}} }
{ \prod_{\ell=k, \ell\neq j}^n [{{\ell}\choose{2}} - {{j}\choose{2}}
] }\qquad
\mbox{for $k<n$ and $k\le j \le n$.}
\]
We have said in principle earlier because the coefficients grow rapidly
and have alternating signs,
which to quote the authors: ``makes the use of this result for samples
of size $n>50$ difficult.''

Fortunately, for our purposes (\ref{firstco}) is enough. From its
derivation and the inequality
$e^{-x} \ge1-x$ we have
\begin{eqnarray*}
P( T_n > t ) & = &\exp\biggl( - \frac{n(n-1)}{2N\gamma} (e^{\gamma t} -
1) \biggr)\\
& \ge&1 - \frac{n(n-1)}{2N\gamma} e^{\gamma t}.
\end{eqnarray*}
The right-hand side is 0 at time $u_n = (1/\gamma) \log( 2N\gamma
/n(n-1))$ so
\begin{eqnarray}\label{ESkO1}
ET_n &\ge&\frac{1}{\gamma} \log\biggl( \frac{2N\gamma}{n(n-1)}\biggr)
- \frac{n(n-1)}{2N\gamma} \int_0^{u_n} e^{\gamma s} \,ds
\nonumber
\\[-8pt]
\\[-8pt]
\nonumber
& \ge&\frac{1}{\gamma} \biggl[ \log\biggl( \frac{2N\gamma}{n(n-1)}\biggr)
- 1 \biggr].
\end{eqnarray}
This is within $O(1)$ of the time $(1/\gamma)\log N$ at which the model
stops making sense, so it follows that
the expected values of $S_k = T_k - T_{k+1}$ are $O(1)$ for $2\le k < n$.

\section{\texorpdfstring{Proofs of the wave $k$ formulas (\protect\ref{tmedeq}) and (\protect\ref{skrecc})}
{Proofs of the wave k formulas (18) and (19)}}\label{sec4}
Our next topic is the waiting time for the first type $k+1$:
\[
P( \tau_{k+1}>t| \mathcal{ F}^k_t )
= \exp\biggl( - \int_0^t Z^*_{k}(s) \,ds \biggr) \approx\exp
(-u_{k+1}V_k e^{\lambda_k t}/\lambda_k).
\]
Taking expected value and using Theorem~\ref{Zklimth}
\[
P( \tau_{k+1} > t | \Omega^0_\infty)
= \bigl( 1 + c_{\theta,k} \mu_{k} (u_{k+1} e^{\lambda_kt}/\lambda
_k)^{\lambda_0/\lambda_k} \bigr)^{-1}.
\]
Using the definition of $\mu_{k+1}$ the median $t^{k+1}_{1/2}$ is
defined by
\[
c_{\theta,k} \mu_{k+1} \exp(\lambda_0 t^{k+1}_{1/2}) \lambda
_k^{-\lambda
_0/\lambda_k} = 1
\]
and solving gives
\[
t^{k+1}_{1/2} = \frac{1}{\lambda_0} \log\biggl(
\frac{ \lambda_k^{\lambda_0/\lambda_k} }{ c_{\theta,k} \mu_{k+1} }
\biggr)
= \frac{1}{\lambda_k} \log(\lambda_k)
- \frac{1}{\lambda_0} \log( c_{\theta,k} \mu_{k+1} )
\]
which is (\ref{tmedeq}). As in the case of $\tau_1$
\[
P( \tau_{k+1} > t^{k+1}_{1/2} + x/\lambda_0 ) \approx( 1 + e^x )^{-1}.
\]

Again the result for the median $s^{k+1}_{1/2}$ of the time $\sigma
_{k+1}$ of the first mutation to type $k+1$ with a family that does not
die out can be found by replacing $u_{k+1}$ by $u_{k+1} \lambda_{k+1}/a_{k+1}$.
Using $\mu_{k+1} = \mu_k u_{k+1}^{\lambda_0/\lambda_k}$ from (\ref
{mukrec}) when we do this gives
\begin{equation}
s^{k+1}_{1/2} = \frac{1}{\lambda_k} \log\biggl( \frac{ \lambda_k a_{k+1}
}{ u_{k+1}\lambda_{k+1} }\biggr)
- \frac{1}{\lambda_0} \log( c_{\theta,k} \mu_{k}).
\label{skmedeq}
\end{equation}

To simplify and to relate our result to (\ref{err2}), we will look at
the difference
\begin{eqnarray*}
s^{k+1}_{1/2} - s^k_{1/2} &=& \frac{1}{\lambda_k} \log\biggl( \frac{\lambda
_k a_{k+1} }{ u_{k+1} \lambda_{k+1} } \biggr)
- \frac{1}{\lambda_{k-1}} \log\biggl( \frac{ \lambda_{k-1} a_{k} }{ u_k
\lambda_k } \biggr)
\\
&&{}- \frac{1}{\lambda_0} \log( c_{h,k}^{\lambda_0/\lambda_{k-1} }
u_k^{\lambda_0/\lambda_{k-1}} ),
\end{eqnarray*}
where in the second term we have used (\ref{cthrec}) and (\ref{mukrec})
to evaluate
$c_{\theta,k}/c_{\theta,k-1}$ and $\mu_k/\mu_{k-1}$. Recalling the formula
\[
c_{h,k} = \frac{1}{a_{k}} \biggl( \frac{a_k}{\lambda_k} \biggr)^{\alpha
_k} \Gamma(\alpha_k)\Gamma(1-\alpha_k)\qquad
\mbox{with } \alpha_k = \lambda_{k-1}/\lambda_k
\]
given in (\ref{chkdef}) we have
\[
s^{k+1}_{1/2} - s^k_{1/2} = \frac{1}{\lambda_k} \log\biggl( \frac{\lambda
^2_k a_{k+1} }{ a_k u_{k+1}\lambda_{k+1} } \biggr)
- \frac{1}{\lambda_{k-1}} \log\bigl( \alpha_k \Gamma(\alpha_k)\Gamma
(1-\alpha
_k) \bigr)
\]
which is (\ref{skrecc}). To see this note that the $u_k$ from the last
term and the $1/a_k$ from the $c_{h,k}$ cancel with parts of the second
term, and the $(a_k/\lambda_k)^{\alpha_k}$ from the third ends up in
the first.

\section*{Acknowledgements}
The author would like to express his appreciation to the AE and
referee, whose many suggestions,
especially their suggested reorganization of the presentation of the
results, greatly improved the paper.


%

\printaddresses


\begin{thebibliography}{26}

\bibitem[\protect\citeauthoryear{Bozic et~al.}{2010}]{Bozetal10}
%
\begin{barticle}[pbm]
\bauthor{\bsnm{Bozic},~\bfnm{Ivana}\binits{I.}},
\bauthor{\bsnm{Antal},~\bfnm{Tibor}\binits{T.}},
\bauthor{\bsnm{Ohtsuki},~\bfnm{Hisashi}\binits{H.}},
\bauthor{\bsnm{Carter},~\bfnm{Hannah}\binits{H.}},
\bauthor{\bsnm{Kim},~\bfnm{Dewey}\binits{D.}} \betal{et al.}
(\byear{2010}).
\btitle{Accumulation of driver and passenger mutations during tumor
progression}.
\bjournal{Proc. Natl. Acad. Sci. USA}
\bvolume{107}
\bpages{18545--18550}.
\bid{doi={10.1073/pnas.1010978107}, issn={1091-6490}, pii={1010978107},
pmcid={2972991}, pmid={20876136}}
\bptok{imsref}%
\end{barticle}
%
\endbibitem

\bibitem[\protect\citeauthoryear{The Cancer Genome Atlas Research
Network}{2008}]{Net08}
%
\begin{bmisc}[pbm]
\borganization{The Cancer Genome Atlas Research Network}
(\byear{2008}).
\bhowpublished{Comprehensive genomic characterization defines human
glioblastoma genes and core pathways. \textit{Nature} \textbf{455}
1061--1068.}
\bid{doi={10.1038/nature07385}, issn={1476-4687}, mid={NIHMS68048},
pii={nature07385}, pmcid={2671642}, pmid={18772890}}
\bptok{imsref}%
\end{bmisc}
%
\endbibitem

\bibitem[\protect\citeauthoryear{Darling}{1952}]{Da52}
%
\begin{barticle}[auto]
\bauthor{\bsnm{Darling},~\bfnm{D.~A.}\binits{D.~A.}}
(\byear{1952}).
\btitle{The role of the maximum term in the sum of independent random variables}.
\bjournal{Trans. Amer. Math. Soc.}
\bvolume{73}
\bpages{95--107}.
\bptok{imsref}%
\end{barticle}
%
\endbibitem

\bibitem[\protect\citeauthoryear{Durrett}{2008}]{Dur}
%
\begin{bbook}[auto:STB|2012/04/12|05:18:16]
\bauthor{\bsnm{Durrett},~\bfnm{R.}\binits{R.}}
(\byear{2008}).
\btitle{Probability Models for DNA Sequence Evolution},
\bedition{2nd} ed.
\bpublisher{Springer}, \baddress{New York}.
\bptok{imsref}%
\end{bbook}
%
\endbibitem

\bibitem[\protect\citeauthoryear{Durrett and Moseley}{2010}]{DurMos10}
%
\begin{barticle}[auto:STB|2012/04/12|05:18:16]
\bauthor{\bsnm{Durrett},~\bfnm{R.}\binits{R.}} \AND
\bauthor{\bsnm{Moseley},~\bfnm{S.}\binits{S.}}
(\byear{2010}).
\btitle{Evolution of resistance and progression to disease during clonal
expansion of cancer}.
\bjournal{Theor. Pop. Biol.}
\bvolume{77}
\bpages{42--48}.
\bptok{imsref}%
\end{barticle}
%
\endbibitem

\bibitem[\protect\citeauthoryear{Durrett and Schweinsberg}{2004}]{DurSch04}
%
\begin{barticle}[auto:STB|2012/04/12|05:18:16]
\bauthor{\bsnm{Durrett},~\bfnm{R.}\binits{R.}} \AND
\bauthor{\bsnm{Schweinsberg},~\bfnm{J.}\binits{J.}}
(\byear{2004}).
\btitle{Approximating selective sweeps}.
\bjournal{Theor. Pop. Biol.}
\bvolume{66}
\bpages{129--138}.
\bptok{imsref}%
\end{barticle}
%
\endbibitem

\bibitem[\protect\citeauthoryear{Durrett and Schweinsberg}{2005}]{DurSch05}
%
\begin{barticle}[auto:STB|2012/04/12|05:18:16]
\bauthor{\bsnm{Durrett},~\bfnm{R.}\binits{R.}} \AND
\bauthor{\bsnm{Schweinsberg},~\bfnm{J.}\binits{J.}}
(\byear{2005}).
\btitle{Power laws for family sizes in a gene duplication model}.
\bjournal{Ann. Probab.}
\bvolume{33}
\bpages{2094--2126}.
\bptok{imsref}%
\end{barticle}
%
\endbibitem


\bibitem[\protect\citeauthoryear{Durrett et~al.}{2011}]{Duretal11}
%
\begin{barticle}[auto:STB|2012/04/12|05:18:16]
\bauthor{\bsnm{Durrett},~\bfnm{R.}\binits{R.}},
\bauthor{\bsnm{Foo},~\bfnm{J.}\binits{J.}},
\bauthor{\bsnm{Ledeer},~\bfnm{K.}\binits{K.}},
\bauthor{\bsnm{Mayberry},~\bfnm{J.}\binits{J.}} \AND
\bauthor{\bsnm{Michor},~\bfnm{F.}\binits{F.}}
(\byear{2011}).
\btitle{Intratumor heterogeneity in evolutionary models of tumor progression}.
\bjournal{Genetics}
\bvolume{188}
\bpages{461--477}.
\bptok{imsref}%
\end{barticle}
%
\endbibitem

\bibitem[\protect\citeauthoryear{Fuchs, Joffe and Teugels}{2001}]{FJT01}
%
\begin{barticle}[auto6]
\bauthor{\bsnm{Fuchs},~\bfnm{A.}\binits{A.}},
\bauthor{\bsnm{Joffe},~\bfnm{A.}\binits{A.}} \AND
\bauthor{\bsnm{Teugels},~\bfnm{J.}\binits{J.}}
(\byear{2001}).
\btitle{Expectation of the ratio of the sum of squares to the square of
the sum: exact and asymptotic results}.
\bjournal{Theory Probab. Appl.}
\bvolume{46}
\bpages{243--255}.
\bptok{imsref}%
\end{barticle}
%
\endbibitem

\bibitem[\protect\citeauthoryear{Griffiths and Pakes}{1988}]{GriPak88}
%
\begin{barticle}[auto:STB|2012/04/12|05:18:16]
\bauthor{\bsnm{Griffiths},~\bfnm{R.~C.}\binits{R.~C.}} \AND
\bauthor{\bsnm{Pakes},~\bfnm{A.~G.}\binits{A.~G.}}
(\byear{1988}).
\btitle{An infinite-alleles version of the simple branching process}.
\bjournal{Adv. in Appl. Probab.}
\bvolume{20}
\bpages{489--524}.
\bptok{imsref}%
\end{barticle}
%
\endbibitem

\bibitem[\protect\citeauthoryear{Griffiths and Tavar{\'e}}{1998}]{GriTav98}
%
\begin{barticle}[auto:STB|2012/04/12|05:18:16]
\bauthor{\bsnm{Griffiths},~\bfnm{R.~C.}\binits{R.~C.}} \AND
\bauthor{\bsnm{Tavar{\'e}},~\bfnm{S.}\binits{S.}}
(\byear{1998}).
\btitle{The age of mutation in the general coalescent tree}.
\bjournal{Stoch. Models}
\bvolume{14}
\bpages{273--295}.
\bptok{imsref}%
\end{barticle}
%
\endbibitem

\bibitem[\protect\citeauthoryear{Haeno, Iwasa and Michor}{2007}]{HaeIwaMic07}
%
\begin{barticle}[pbm]
\bauthor{\bsnm{Haeno},~\bfnm{Hiroshi}\binits{H.}},
\bauthor{\bsnm{Iwasa},~\bfnm{Yoh}\binits{Y.}} \AND
\bauthor{\bsnm{Michor},~\bfnm{Franziska}\binits{F.}}
(\byear{2007}).
\btitle{The evolution of two mutations during clonal expansion}.
\bjournal{Genetics}
\bvolume{177}
\bpages{2209--2221}.
\bid{doi={10.1534/genetics.107.078915}, issn={0016-6731}, pii={177/4/2209},
pmcid={2219486}, pmid={18073428}}
\bptok{imsref}%
\end{barticle}
%
\endbibitem

\bibitem[\protect\citeauthoryear{Iwasa, Nowak and Michor}{2006}]{IwaNowMic06}
%
\begin{barticle}[pbm]
\bauthor{\bsnm{Iwasa},~\bfnm{Yoh}\binits{Y.}},
\bauthor{\bsnm{Nowak},~\bfnm{Martin~A.}\binits{M.~A.}} \AND
\bauthor{\bsnm{Michor},~\bfnm{Franziska}\binits{F.}}
(\byear{2006}).
\btitle{Evolution of resistance during clonal expansion}.
\bjournal{Genetics}
\bvolume{172}
\bpages{2557--2566}.
\bid{doi={10.1534/genetics.105.049791}, issn={0016-6731},
pii={genetics.105.049791}, pmcid={1456382}, pmid={16636113}}
\bptok{imsref}%
\end{barticle}
%
\endbibitem

\bibitem[\protect\citeauthoryear{Jones et al.}{2008}]{Jon08}
%
\begin{barticle}[auto:STB|2012/04/12|05:18:16]
\bauthor{\bsnm{Jones},~\bfnm{S.}\binits{S.}} \betal{et al.}
(\byear{2008}).
\btitle{Core signalling pathways in human pancreatic cancers revealed
by global
genomic analyses}.
\bjournal{Science}
\bvolume{321}
\bpages{1801--1812}.
\bptok{imsref}%
\end{barticle}
%
\endbibitem

\bibitem[\protect\citeauthoryear{Jones et al.}{2010}]{Jon10}
%
\begin{barticle}[auto:STB|2012/04/12|05:18:16]
\bauthor{\bsnm{Jones},~\bfnm{S.}\binits{S.}} \betal{et al.}
(\byear{2010}).
\btitle{Frequent mutations of chromatic remodeling gene ARID1A in
ovarian cell
carcinoma}.
\bjournal{Science}
\bvolume{330}
\bpages{228--231}.
\bptok{imsref}%
\end{barticle}
%
\endbibitem

\bibitem[\protect\citeauthoryear{Kingman}{1982}]{Kin}
%
\begin{bincollection}[auto:STB|2012/04/12|05:18:16]
\bauthor{\bsnm{Kingman},~\bfnm{J.~F.~C.}\binits{J.~F.~C.}}
(\byear{1982}).
\btitle{Exchangeability and the evolution of large populations}.
In \bbooktitle{Exchangeability in Probability and Statistics}
(\beditor{\binits{G.} \bsnm{Koch}}
\AND
\beditor{\binits{F.} \bsnm{Spizzechio}}, eds.)
\bpages{97--112}.
\bpublisher{North-Holland}, \baddress{Amsterdam}.
\bptok{imsref}
\end{bincollection}
\endbibitem

\bibitem[\protect\citeauthoryear{Logan et al.}{1973}]{Lo73}
%
\begin{barticle}[auto]
\bauthor{\bsnm{Logan},~\bfnm{B.~F.}\binits{B.~F.}},
\bauthor{\bsnm{Mallows},~\bfnm{C.~L.}\binits{C.~L.}},
\bauthor{\bsnm{Rice},~\bfnm{S.~O.}\binits{S.~O.}}
\AND
\bauthor{\bsnm{Shepp},~\bfnm{L.~A.}\binits{L.~A.}}
(\byear{1973}).
\btitle{Limit distributions of self-normalized sums}.
\bjournal{Ann. Probab.}
\bvolume{1}
\bpages{788--809}.
\bptok{imsref}%
\end{barticle}
%
\endbibitem


\bibitem[\protect\citeauthoryear{Luebeck and Mollgavkar}{2002}]{LueMol02}
%
\begin{barticle}[auto:STB|2012/04/12|05:18:16]
\bauthor{\bsnm{Luebeck},~\bfnm{E.~G.}\binits{E.~G.}} \AND
\bauthor{\bsnm{Mollgavkar},~\bfnm{S.~H.}\binits{S.~H.}}
(\byear{2002}).
\btitle{Multistage carcinogenesis and the incidence of colorectal cancer}.
\bjournal{Proc. Natl. Acad. Sci. USA}
\bvolume{99}
\bpages{15095--15100}.
\bptok{imsref}%
\end{barticle}
%
\endbibitem

\bibitem[\protect\citeauthoryear{O'Connell}{1993}]{OCo93}
%
\begin{barticle}[auto:STB|2012/04/12|05:18:16]
\bauthor{\bsnm{O'Connell},~\bfnm{N.}\binits{N.}}
(\byear{1993}).
\btitle{Yule approximation for the skeleton of a branching process}.
\bjournal{J.~Appl. Probab.}
\bvolume{30}
\bpages{725--729}.
\bptok{imsref}%
\end{barticle}
%
\endbibitem

\bibitem[\protect\citeauthoryear{Parmigiani et al.}{2007}]{Par}
%
\begin{bmisc}[auto:STB|2012/04/12|05:18:16]
\bauthor{\bsnm{Parmigiani},~\bfnm{G.}\binits{G.}} \betal{et al.}
(\byear{2007}).
\bhowpublished{Statistical methods for the analysis of cancer genome seqeuncing
data. Available at \url{http://www.bepress.com/jhubiostat/paper126}}.
\bptok{imsref}%
\end{bmisc}
%
\endbibitem

\bibitem[\protect\citeauthoryear{Parsons et al.}{2008}]{Par08}
%
\begin{barticle}[auto:STB|2012/04/12|05:18:16]
\bauthor{\bsnm{Parsons},~\bfnm{D.~W.}\binits{D.~W.}} \betal{et al.}
(\byear{2008}).
\btitle{An integrated genomic analysis of human glioblastome multiforme}.
\bjournal{Science}
\bvolume{321}
\bpages{1807--1812}.
\bptok{imsref}%
\end{barticle}
%
\endbibitem

\bibitem[\protect\citeauthoryear{Pitman}{2006}]{Pit06}
%
\begin{bbook}[auto:STB|2012/04/12|05:18:16]
\bauthor{\bsnm{Pitman},~\bfnm{J.}\binits{J.}}
(\byear{2006}).
\btitle{Combinatorial Stochastic Processes}.
\bpublisher{Springer}, \baddress{New York}.
\bptok{imsref}%
\end{bbook}
%
\endbibitem

\bibitem[\protect\citeauthoryear{Pitman and Yor}{1997}]{PitYor97}
%
\begin{barticle}[auto:STB|2012/04/12|05:18:16]
\bauthor{\bsnm{Pitman},~\bfnm{J.}\binits{J.}} \AND
\bauthor{\bsnm{Yor},~\bfnm{M.}\binits{M.}}
(\byear{1997}).
\btitle{The two-parameter Poisson--Dirichlet distribution derived from
a stable
subordinator}.
\bjournal{Ann. Probab.}
\bvolume{25}
\bpages{855--900}.
\bptok{imsref}%
\end{barticle}
%
\endbibitem

\bibitem[\protect\citeauthoryear{Polanski, Bobrowski and Kimmel}{2003}]{PolBobKim03}
%
\begin{barticle}[auto:STB|2012/04/12|05:18:16]
\bauthor{\bsnm{Polanski},~\bfnm{A.}\binits{A.}},
\bauthor{\bsnm{Bobrowski},~\bfnm{A.}\binits{A.}} \AND
\bauthor{\bsnm{Kimmel},~\bfnm{M.}\binits{M.}}
(\byear{2003}).
\btitle{A note on distributions of times to coalescence, under time-dependent
population size}.
\bjournal{Theor. Pop. Biol.}
\bvolume{63}
\bpages{33--40}.
\bptok{imsref}%
\end{barticle}
%
\endbibitem

\bibitem[\protect\citeauthoryear{Sj{\"{o}}blom et~al.}{2006}]{Sjoetal06}
%
\begin{barticle}[pbm]
\bauthor{\bsnm{Sj{\"{o}}blom},~\bfnm{Tobias}\binits{T.}} \betal{et al.}
(\byear{2006}).
\btitle{The consensus coding sequences of human breast and colorectal cancers}.
\bjournal{Science}
\bvolume{314}
\bpages{268--274}.
\bid{doi={10.1126/science.1133427}, issn={1095-9203}, pii={1133427},
pmid={16959974}}
\bptok{imsref}%
\end{barticle}
%
\endbibitem

\bibitem[\protect\citeauthoryear{Slatkin and Hudson}{1991}]{SlaHud91}
%
\begin{barticle}[pbm]
\bauthor{\bsnm{Slatkin},~\bfnm{M.}\binits{M.}} \AND
\bauthor{\bsnm{Hudson},~\bfnm{R.~R.}\binits{R.~R.}}
(\byear{1991}).
\btitle{Pairwise comparisons of mitochondrial DNA sequences in stable and
exponentially growing populations}.
\bjournal{Genetics}
\bvolume{129}
\bpages{555--562}.
\bid{issn={0016-6731}, pmcid={1204643}, pmid={1743491}}
\bptok{imsref}%
\end{barticle}
%
\endbibitem

\bibitem[\protect\citeauthoryear{Wood et al.}{2007}]{Woo07}
%
\begin{barticle}[auto:STB|2012/04/12|05:18:16]
\bauthor{\bsnm{Wood},~\bfnm{L.~D.}\binits{L.~D.}} \betal{et al.}
(\byear{2007}).
\btitle{Tyhe genomic landscapes of human breast and colorectal cancers}.
\bjournal{Science}
\bvolume{318}
\bpages{1108--1113}.
\bptok{imsref}%
\end{barticle}
%
\endbibitem

\end{thebibliography}
\end{document}